\documentclass[11pt,twoside]{article}
\usepackage{asp2010}

\resetcounters

\begin{document}

\title{Globular cluster systems of early-type galaxies - do we  understand them?}
\author{Tom Richtler
\affil{Departamento de Astronom\'{\i}a, Universidad de Concepci\'on, Barrio Universitario, Concepci\'on, Chile}}

\begin{abstract}
I review recent and less recent work on  globular cluster systems in early-type galaxies.  Explaining their properties
 and possible  assembly
scenarios, touches on a variety of astrophysical topics from  cluster formation itself to galaxy formation and evolution and even details of observational techniques. 
The spectacular  cluster systems of central galaxies in galaxy clusters may owe their richness to a plethora of less spectacular galaxies and their
star formation processes. It seems that dwarf galaxies occupy a particularly important role.  
\end{abstract}

\section{Introduction -  globular clusters and dwarf galaxies}

Talking about globular cluster systems with the intention to reach
an audience of  non-experts is a particular challenge because often, a specific
discovery or result is sparkling brighter inside the cosmos of experts than outside, where the  horizon is wider
and sparks dim easily.  On the other hand, the variety of astrophysical problems relevant to star cluster research (see the reviews of \citealt{brodie06}, \citealt{harris10}, and do not miss \citealt{kissler09})   
joins a large and  explosive congregation, where sparks can have tremendous effects.

The formation of globular clusters (GCs) is yet to be understood in detail (as  is usually the case with dissipative processes in astrophysics), but it is not a mystery. We see GCs today forming in a variety of
galaxy types, most spectacularly in starbursts, triggered by galaxy interactions (e.g. \citealt{schweizer09}).    The origin of the most
massive star cluster known, with a mass of  $\mathrm 8 \times 10^7 M_\odot$ \citep{maraston04}, W3 in the interacting galaxy NGC 7252, can be convincingly related to the starburst, which
occurred 0.5 Gyr ago. Such massive objects may form by rapid merging of star cluster complexes \citep{fellhauer05,kissler06}. Also in
"normal" spiral galaxies, the formation of massive star clusters is supported by a high star-formation rate \citep{larsen00}. 
Integrating the  cosmic star formation rate (e.g. \citealt{hopkins04}), one finds that about 65\% of all stellar mass has been formed before z=1 (age 8.8 Gyr
for a standard cosmology). In this epoch, the star formation rate was much higher than it is today, so most  GCs in the Universe are  old,
but do not represent by nature  the oldest populations. This "cosmic" argument does not apply to individual galaxies and many intermediate-age GCs have been found in early-type galaxies (e.g. \citealt{puzia05}).\\

Messier 92  is  one of the best investigated Galactic GCs. Its age can currently be constrained
to be $11 \pm 1.5$ Gyr \citep{dicecco10}
($1.6 < z < 5$), and it is one of the metal-poorest clusters. An iron abundance of [Fe/H]=-2.3 \citep{kraft03}
and a mass of $1.5\times10^5$ $M_\odot$ mean a total iron mass of only 0.8$M_\odot$, a mass, which can be
produced by a few SNII supernovae. That there is no detectable star-to-star variation of
the iron abundance (which is not the case for other elements, e.g. \citealt{angelou12}), permits
the conclusion that M92 is not self-enriched and that it has  formed in an already well-mixed environment. This environment
cannot have been very massive since there is no significant field population with this
iron abundance. If we call it a "dwarf galaxy" then we are  close to the scenario proposed
by \citet{searle78} who called these entities "protogalactic fragments".

Dwarf galaxies donate GCs to  the
Milky Way. This becomes manifest  through the Sagittarius stream \citep{ibata01,siegel11} and other candidates (e.g. \citealt{pawlowski12} and
references therein).
Moreover, the  "young halo clusters" trace the plane of Milky Way satellites \citep{yoon02,kroupa05,keller12}. \citet{forbes10}
estimate that an appreciable fraction  of the Galactic GC system has been accreted through dwarf galaxies.

If accretion plays an important role for the assembly of the metal-poor part  of relatively isolated
spiral galaxies, what role does it occupy in really dense environments?


\section{The richness of globular cluster systems}
\subsection{The richest globular cluster systems are not so rich}

A popular quantitative measure for the richness of a globular cluster system (GCS) is the "specific frequency" $S_N$, which  has been  defined by \citep{harris81} as $S_N = N_{GC} 10^{0.4 (M_V+15)}$, where $N_{GC}$ is the total number of GCs and
$M_V$ the host galaxy's absolute V-magnitude.  

 We find the richest GCSs in terms of GC number around
central galaxies in galaxy clusters, the nearest being M87 (Virgo) \citep{tamura06,harris09c,strader11}, NGC 1399 (Fornax) \citep{kissler99,richtler04,dirsch03a,schuberth10}, NGC 3311
 (Hydra I) \citep{wehner08,richtler11}. $S_N$-values for these galaxies, which can host more than 10000 GCs, are somewhat uncertain, not so much
 for the number of GCs, but because $M_V$ can be easily underestimated for these galaxies with very extended stellar halos. 
 The case of NGC 3311 is illustrative, because this galaxy had the reputation of showing a particularly high $S_N$, e.g. \citet{mclaughlin95} quoted $S_N = 15 \pm 6$.
 \citet{wehner08} found 16500 GCs within a radius of 150 kpc and adopted $M_V= -22.8$, thus $S_N = 12.5$, the uncertainties still permitting an extreme  lower limit of $S_N \approx 9$. But integrating the V-luminosity profile of \citet{richtler11} out to the same radius results in $M_V = -24$ and $S_N = 4.1$, which is a normal value for
 giant elliptical galaxies.  A similar point has been made
 for NGC 1399 \citep{ostrov98,dirsch03a}. Therefore there is no hard evidence that $S_N$-values for central giant ellipticals are dramatically higher than for normal ellipticals.

  Intuition tells us that accretion of dwarf galaxies for ellipticals in galaxy clusters should be even more important than for spiral galaxies 
  (\citealt{cote98} formulated this beyond intuition; see also \citealt{hilker99}).
   Clear evidence, e.g. in the form
of streams, is only just now emerging, for example in the cases of M87 \citep{romanowski12} or NGC 3311 \citep{arnaboldi12}. The halos of these central
galaxies have been built up by long-term accretion of material from the cluster environment, a process which is still on-going.
 Strong evidence for a significant growth of massive elliptical galaxies since z=2 (10.3 Gyr) has been provided by \citet{vandokkum10}. They
find that outside a core  with a size of about 5kpc, elliptical galaxies increased their mass by a factor of 4 within the last 10 Gyr. This happens predominantly through minor and dry mergers \citep{tal12}. 
Obviously, GCSs should share the same fate. 

 For NGC 1399, kinematical data indicate that many GCs cannot have  formed {\it in situ}.
 One finds in the GCS of NGC 1399 objects, which by their
high radial velocities must reach apocentric distances of 500  kpc or even greater \citep{richtler04,schuberth10}. These few objects near their pericenters
must trace
a much larger (unknown) population with high space velocities, but low radial velocities. 
Their orbital velocities
result from  potential differences that exist within the Fornax cluster rather than within NGC 1399, so one may call them "intracluster GCs"  
\citep{kissler99}.

\subsection{The poorest globular cluster systems can be the richest}
\label{sec:dwarfs}
The GCSs of early-type dwarf galaxies are as interesting as those of giant ellipticals. The highest specific
frequency known is that of the Fornax dwarf spheroidal with 5 GCs ($S_N\approx30$). Are dwarf galaxies for some
reason more efficient in
forming GCs?  \citet{miller07} indeed found a trend of increasing $S_N$  with decreasing brightness of  the host galaxy  for dwarf ellipticals in Fornax,
Virgo and the Leo group. The largest
data base in this respect is the HST/ACS Virgo survey \citep{peng08}, in which about 100 early-type Virgo galaxies have been imaged down
to a magnitude of  $M_V \approx -16$. This work does not support a clear relation between $S_N$ and host galaxy brightness, but dwarf
galaxies fill a larger $S_N$-interval than giant ellipticals. Probably a key finding is that dwarf galaxies
with large clustercentric distances consistently show low $S_N$-values, while high  $S_N$-values are  found among the (projected) inner dwarf galaxy
population, i.e. among those galaxies  with a higher probability of interactions, which may have triggered star-bursts.

\section{The phenomenon of "bimodality" }
The metallicity distribution of Galactic GCs is "bimodal", i.e. well represented by two Gaussians, with metal-rich and metal-poor
GCs being the bulge and the halo clusters, respectively \citep{zinn85}. Do we find a similar fundamental structure in the GCSs of  elliptical galaxies?

In an influential paper, \citet{ashman92} explored the idea that GCs form in mergers
and interactions of galaxies (see their introduction for the history of this
concept) and hypothesized that the metallicity distribution of GCs in elliptical
galaxies should be bimodal. Adopting the merger paradigm for
elliptical galaxies, metal-poor clusters are the old GCs of the pre-merger components,
while metal-rich clusters form in starbursts triggered by gas-rich mergers of spiral galaxies. 
This {\it prediction} of bimodality in the {\it color} distribution of GCs has indeed been found in many elliptical galaxies (e.g. \citealt{zepf93,whitmore95,gebhardt99,larsen01,kundu01,peng06,harris09a,harris09b}). Studies in the Washington photometric system showed this
bimodality particularly well (e.g. \citealt{geisler96, dirsch03a,dirsch03b,bassino06}): it can be
characterized by two Gaussians with peaks at C-R=1.35 (the "blue" peak) and  C-R=1.75 (the "red" peak) (these peaks are not found in Fig.\ref{fig:n1316_utrecht}!). The blue peak is remarkably constant among the investigated galaxies,
 while the red peak
gets
slightly redder with increasing  host galaxy brightness (see also \citealt{larsen00}). However, this color bimodality does not
apply to the brightest clusters in some GCSs, which avoid very red and very blue colors.

  The HST/ACS surveys  in  Virgo and the Fornax  confirm this with a much larger database 
\citep{peng06,peng08,jordan09}. 
It  turns out that bimodal
color distributions are mainly a signature of bright host galaxies, which does not come as a surprise: early-type galaxies follow a well-defined color-magnitude
relation (e.g. \citealt{smithcastelli08,misgeld08}) and one does not expect GCs to be redder than the host galaxy itself.
 A  striking difference between the blue and red clusters is that the mean color of the red subpopulation strongly correlates with  the host galaxy luminosity and thus with its color, while the mean color of blue clusters is more or less constant.
  
Blue and red GC populations, however, differ in more than only their colors. As shown in many papers, the radial number density profiles of red clusters are 
steeper and resemble more the light profile of their host galaxies. Accordingly, they show a lower velocity dispersion than the blue clusters.  For NGC 1399, Fig.20 of \citet{schuberth10} demonstrates that 
the velocity dispersion exhibits a sudden change between red and blue clusters, not a smooth transition. 


 
\begin{figure}[h]
\begin{center}
\includegraphics[width=0.7\textwidth]{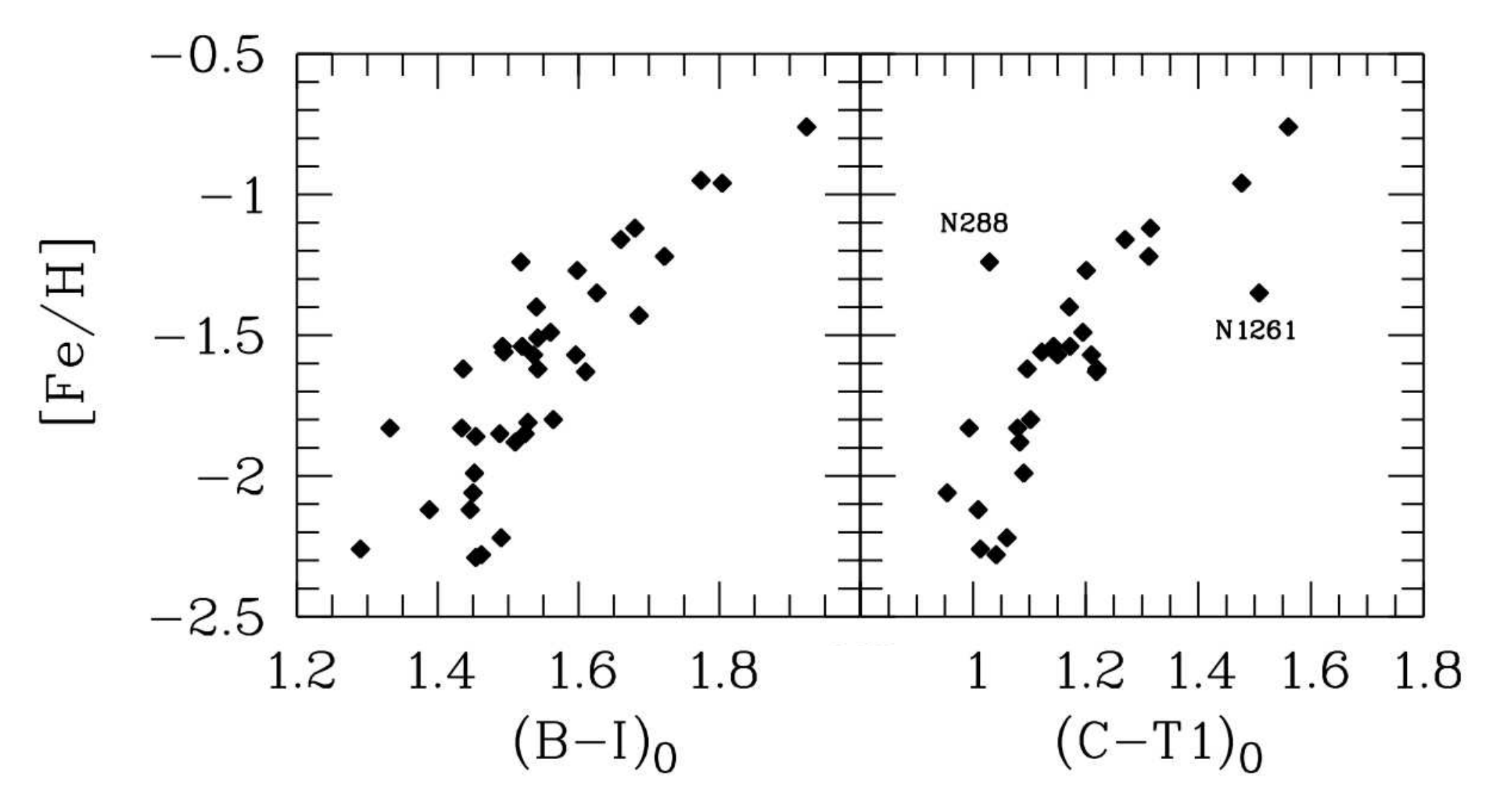}
\caption{Colour-metallicity relations for Galactic GCs with reddenings less than E(B-V)=0.15, using B-I data from \citet{harris96} and C-T1 data from \citet{harris77}.  The non-linearity is clearly seen.  For
clusters metal-poorer than about [Fe/H]=-1.5,  the color becomes a very bad proxy for the metallicity, more so for B-I than for C-T1. Note, however, the large deviations for some clusters. Also note
that these colors, measured with diaphragms, refer only to the innermost parts of GCs. }
\label{fig:BIcolor}
\end{center}
\end{figure}

Do these bimodal color distributions reflect bimodal 
metallicity distributions?   
  There had been some claims that a non-linearity of the color-metallicity relation, in combination with
a considerable scatter around this relation, can produce a bimodal color distribution, even if the
underlying metallicity distribution is not bimodal \citep{richtler06,yoon06}.  More recent work  strengthens this point. \citet{yoon11a,yoon11b} derive
GC metallicity distributions from a non-linear color-metallicity relation and show that the inferred metallicity distributions are rather single-peaked
(I caution that their example NGC 4374 is a multiple SNIa host galaxy, see more remarks below).
A further complicating point is that their color-metallicity relation becomes so insensitive to metallicity for clusters metal-poorer than about [Fe/H]=-1.5, that to infer a metallicity distribution from
a color distribution seems to be only meaningful if color and metallicity are related without any generic scatter. But this is not the case, because metal-poor Galactic GCs having the same metallicity  can show quite different CMDs (Fig.\ref{fig:BIcolor}).
Therefore, the  reconstructed metallicity distribution for metal-poor clusters might be considered as a mean distribution without saying much
about an individual object.
However, a weak metallicity-brightness relation for metal-poor clusters, as described by \citet{harris09b},  is expected, because massive
clusters have a higher probability to originate from more massive host galaxies with an overall higher metallicity. 

If IR-bands are included, bimodality may vanish almost completely, as shown by \citet{blakeslee12} for NGC 1399, and  \citet{chies11,chies12}  for a sample of  17 early-type galaxies.
Galaxies with very pronounced 
blue peaks in g-z are NGC 4374 and NGC 4526. These galaxies hosted three and two SNIa events, respectively. Therefore, one may assume 
significant intermediate-age populations to be present, and probably also intermediate-age GCs, and the color is not anymore a pure metallicity indicator. More SNIa host galaxies in their sample 
are  NGC 4382, NGC 4621, and NGC 4649.
The striking deficiency of red clusters in NGC 4660 (which one also finds  in the  sample of \citet{peng06}) is remarkable. There may be more individuality among GCSs than previously thought.
In those cases, where metallicities of larger GC samples have been derived from integrated spectra (not only of early-type galaxies!) \citep{foster10,foster11,caldwell11},
the metallicity distributions appear unimodal with one broad peak around [Fe/H]=-1.

\section{Once more: giant and  dwarf galaxies}
The question to what degree $S_N$-values of galaxies reflect the efficiency of GC formation is difficult to answer  (relating GC masses 
to host galaxy mass  as do e.g. \citealt{peng08,georgiev10} would be more physical, but observationally more difficult to determine). 
Although the  $S_N$-values of central galaxies may not be as high as previously thought,  they are still higher than elliptical galaxies of lower 
luminosity.
The ACS Virgo survey \citep{peng08} reveals a shallow
minimum around $M_V \approx$ -20, but without a well-defined  relation  for the dwarf or for the giant regime, although many dwarf galaxies  as faint as $M_V = -16$ 
show $S_N$-values
rivalling or exceeding typical values for giant galaxies.  Including fainter galaxies from \citet{lotz04} lets the trend of increasing
$S_N$ with decreasing luminosity appears clearer: values higher than 10 are normal, particularly for nucleated dwarf galaxies. 

This is also visible in the compilation of \citet{georgiev10}. They   embed the variation of $S_N$-values in the context of galaxy formation, inspired
by \citet{dekel06} (see  \citealt{forbes05} for an earlier account). In brief,  star formation in low mass galaxies is regulated by stellar feedback, in high mass galaxies by virial shocks, which in  both
regimes leads to a suppression of field star formation, and favors star clusters. One notes that this interpretation implies an alternative
view: here it is a generic property of star formation processes in giant ellipticals, which produces the rich GCSs, not the infall of less massive
galaxies with intrinsically higher $S_N$-values. It may be, however, difficult to defend this view in front of all evidence for the role of accretion. 
 Starbursts in dwarf galaxies might hold the key for a proper understanding. 



Can metallicity itself be a parameter for efficient GC formation?  At low metallicity, the energy and momentum input into the interstellar medium via
radiation-driven stellar winds is reduced \citep{kudritzki02}, 
leading
to a higher star formation efficiency \citep{dib11}. \citet{glover12a} find that molecules are not necessarily a prerequisite for star formation, but that dust
is an important ingredient for cooling processes. Star formation occurs in cold gas at higher temperatures, and Jeans masses are increased \citep{glover12b}.
The formation of dense substructures in a collapsing cloud is efficiently suppressed at low metallicities,  because turbulence cannot create clumps
as efficiently as in high-metallicity clouds \citep{glover12b}. This may lead to a more coherent star formation inside a star-forming cloud. 
These are all factors which  support  the dynamical stability of star-forming clouds, and plausibly favor the formation of compact and  coherent  structures,
which lead to the extremely clustered star formation for example in Blue Compact Galaxies (e.g. \citealt{adamo11}).
  
\section{Globular clusters outside galaxies}
"Cosmological" formation of GCs, i.e. GCs embedded in a dark halo, is interesting to imagine, but difficult to prove.  GCs apparently exist outside
galaxies, however, this  does not mean that they were formed outside galaxies.  The objects with high radial velocities around central galaxies spend most of
their orbital life far way from the cluster center. But there are also GC populations in galaxy clusters without a central galaxy.  \citet{west11} found in
 a population of intergalactic GCs in Abell 1185, mostly metal-poor. A recent survey with HST/ACS of the Coma cluster uncovered a huge population of GCs, filling the
entire cluster core  \citep{peng11}. The authors estimate a total of about 47000 GCs out to a radius of 570 kpc.  Dissolution of dwarf galaxies and tidal stripping from 
more massive galaxies might both contribute  to create this largest GCS in the local Universe.

\section{Isolated elliptical galaxies}
Suspecting dwarf galaxies as donators of metal-poor GCs, it would be interesting to compare cluster
galaxies with
isolated elliptical galaxies.  Unfortunately, the data are sparse. After the compilation
of \citet{spitler08}, no new work on isolated ellipticals has been reported,
leaving  NGC 720 \citep{kissler96} and NGC 821 \citep{spitler08} as prototypes. Both
present relatively poor cluster systems. However, many "isolated ellipticals" exhibit
tidally disturbed features and may be in
fact late mergers \citep{tal09}, including  NGC 720 that does not exhibit  obvious morphological peculiarities, but strong population gradients \citep{rembold05}.




\section{NGC 1316 -  cluster formation in a late merger} 
\begin{figure}[h]
\begin{center}
\includegraphics[width=0.8\textwidth]{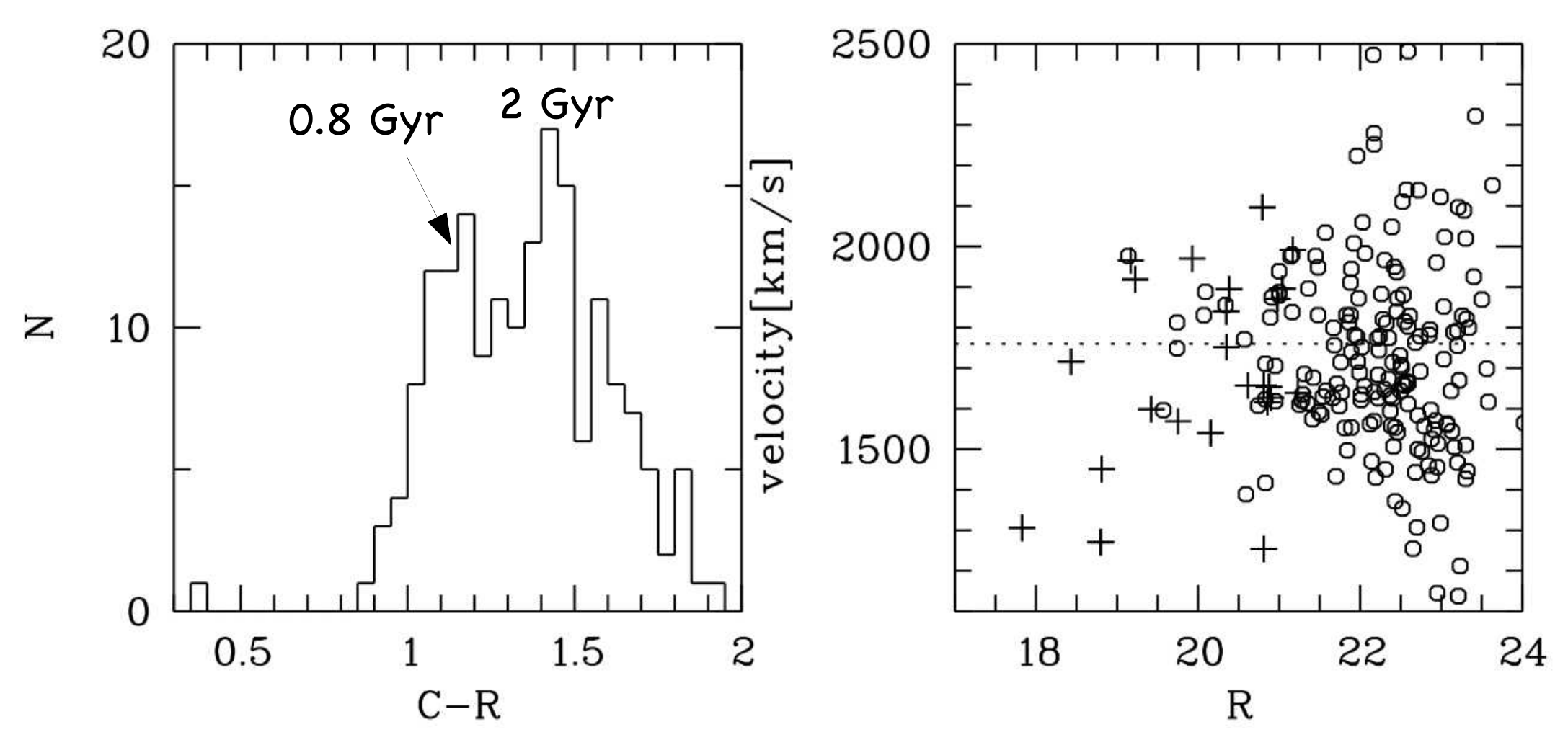}
\caption{ Left panel: color histogram  of 178 confirmed GCs in NGC 1316 (Richtler et al., in preparation). The ages assume solar metallicity. Right panel: radial velocities vs. R-magnitudes for the same clusters augmented by a
sample from \citet{goudfrooij01b} (crosses). Note, how the bright cluster population avoids the systemic velocity (dashed line) and how the velocity dispersion
increases with fainter magnitudes.}  
\label{fig:n1316_utrecht}
\end{center}
\end{figure}

 A galaxy  whose brightness is dominated by intermediate-age populations \citep{kuntschner00}, is the merger remnant NGC 1316 (Fornax A) in the outskirts of the Fornax galaxy cluster. It  might illustrate the processes that were in
  action during the youth of giant ellipticals to form metal-rich clusters (see \citealt{richtler12a} for references).
  It experienced a major starburst about 2-3 Gyr ago that 
produced many massive star clusters, the brightest one (114 in the list of \citealt{goudfrooij01a}) having a mass of the order $2\times10^7 M_\odot$.
Note that \citet{brodie11} would not call it an "Ultra Compact Dwarf", because its effective radius is smaller than 10 pc \citep{goudfrooij12}.
The imprint of the 2 Gyr starburst is a well defined peak in the color distribution of GCs \citep{richtler12a} (Fig.\ref{fig:n1316_utrecht}). Cluster ages from integrated spectra have been
determined only
for a few of the brightest clusters \citep{goudfrooij01b}, but the colors (with the assumption of solar metallicity for all bright clusters) fit well to these ages.  
 Star formation continued after this starburst and a second peak corresponds to 0.8 Gyr (which still has to be confirmed by  spectroscopy). We find GCs with ages as young as 0.5 Gyr. Including fainter clusters probably samples older objects and lets these peaks largely disappear. A very interesting observation is  that the brightest clusters seem to avoid the systemic radial velocity of NGC 1316 by showing large negative offsets up to
500 km/s.  This indicates elongated orbits  and a population of massive clusters far away from the center, which still has to be identified. The starburst seem to have happened in a very early stage of the merger with the
merger components still separated with high relative velocities.

Very remarkable is the object SH2, discovered by \citet{schweizer80}. Rather than as a normal HII-region, it appears as an ensemble of star clusters with
ages around 0.1 Gyr \citep{richtler12b}.  Its exact nature still awaits investigation, but a plausible hypothesis is that of an infalling dwarf galaxy, having
recently experienced a starburst.  In this case, one would expect low metallicities and it would constitute an example, how metal-poor clusters are 
donated to a GCS.  

\section{Inventory}
A cautiously positive answer to the title question seems to be appropriate.
A description of a global picture  (which generously ignores details) can be: giant elliptical galaxies assembled their metal-poor GCs by the accretion of dwarf galaxies. 
 The metal-rich GCs  formed in early starbursts, triggered by gas-rich mergers. Whether  colors alone can provide an 
 adequate description of the substructure of an GCS, has become doubtful.
 With higher precision of kinematical data, more substructures in GCSs will be detected and perhaps single merger events can be identified.
The key to all that is the physics of star cluster formation in starbursts, whether in dwarf galaxies or in massive systems.  Modern 
simulations  of galaxy mergers now resolve star cluster scales (see \citealt{hopkins12} and references therein)  and will probably provide the physical
understanding.    

\acknowledgements{My cordial thanks go to the organizers of this great conference, during which I learned once again to admire the astronomy 
coming out of 
Utrecht.  I thank Soeren S. Larsen for helpful remarks on the present article and for much more.  I am grateful to Richard Lane for
a critical reading of the text. I also acknowledge the financial support  from  the Chilean Center for Astrophysics,
FONDAP Nr. 15010003,   FONDECYT project Nr. 1100620, and
through the BASAL Centro de Astrofisica y Tecnologias
Afines (CATA) PFB-06/2007.  }

\bibliography{utrecht}
\end{document}